# A conducting nano-filament (CNF) network as a precursor to the origin of superconductivity in electron-doped copper oxides

Heshan Yu[1], Ge He[1], Ziquan Lin[2], Anna Kusmartseva[4], Jie Yuan[1], Beiyi Zhu[1], Yi-feng Yang[1,3], Tao Xiang[1,3], Liang Li[2], Junfeng Wang[2,†], F. V. Kusmartsev[4,†] and Kui Jin[1,3,†]

[1]Beijing National Laboratory for Condensed Matter Physics, Institute of Physics, Chinese Academy of Sciences, Beijing 100190, China

[2]Wuhan National High Magnetic Field Center (WHMFC), Huazhong University of Science and Technology, Wuhan 430074, China

[3]Collaborative Innovation Center of Quantum Matter, Beijing, 100190, China

[4]Department of Physics, Loughborough University, Loughborough LE11 3TU, United Kingdom

**Emergence of superconductivity at the instabilities of antiferromagnetism (AFM), spin/charge density waves has been widely recognized in unconventional superconductors [1–3]. Notwithstanding the predominant role played by the spin fluctuations in the superconductivity of electron-doped cuprates [4] the existence of an AFM critical end point is still in controversy for different probes.[5] Here, by tuning the oxygen content, a systematic study of the Hall signal and magnetoresistivity up to 58 Tesla on optimally doped $La_{2-x}Ce_xCuO_{4±\delta}$ (*x* = 0.10) thin films identifies two characteristic temperatures at 62.5 ± 7.5 K and 25.0 ± 5 K. The former is quite robust, whereas the latter becomes flexible in magnetic field, thereby linking respectively to two- and three-dimensional AFM, evident from the published multidimensional phase diagram.[6, 7] This extraordinary observation of an extended antiferromagnetic (AF) phase in contrast to muon spin resonance ($\mu$SR) reports[8] together with a rigorous theoretical analysis of the presented data suggest the existence of conductive nano-filamentary (CNF) structures that effectively corroborate all previously reported field studies. The new findings provide a uniquely consistent alternative picture in understanding the interactions between AFM and superconductivity in electron-doped cuprates and offer a consolidating interpretation to the pioneering scaling law in cuprates recently established by Bozevic et al. [52]**

Copper oxide superconductors (cuprates) demonstrate multiple competing orders, the proximity of which hosts the superconducting phase [9]. Significant efforts have been made to explain these effects [10]. Diverse advanced structural and transport probes have been utilized to determine the critical doping levels for correlative Fermi surface reconstruction, quantum critical points, and various symmetry breaking phenomena. [5, 11, 12] However, the precise position of the critical point is still in controversy, obstructing the approach to the origin of high-Tc superconductivity. The AFM fades away more gradually in electron-doped cuprates than in their hole-doped counterparts,[4] making them the ideal starting point to determine the boundary of the AFM and uncover any key role it has to play in the nature of superconductivity.

For electron-doped $Pr_{2-x}Ce_xCuO_4$ and $Nd_{2-x}Ce_xCuO_4$, transport probes present a roughly consistent critical point scenario at ($x_{FS} \sim 0.16$), evidenced by numerous transport probes [13,14,15,16], a dramatic change in Shubnikov-de Haas quantum oscillations[17], angle-resolved photoemission spectroscopy studies [18,19] and optical spectroscopy[20]. At the same time, neutron scattering experiments[21] show absence of long range AF order above $x \sim 0.13$, just prior to the appearance of the superconductivity.

Similar discrepancy was also observed in $La_{2-x}Ce_xCuO_{4\pm\delta}$ (LCCO), a unique system investigated over the complete phase diagram[6] and demonstrating optimal doping at $x \sim 0.10$. Previous transport measurements reveal the critical doping range between $0.13 < x_{FS} < 0.15$, based on electrical resistivity[22], the Hall coefficient[23] and the in-plane angular dependent magnetoresistivity[24]. However, $\mu$SR probe only sees long-range AFM until a doping level of $x \sim 0.08$, in proximity to the superconducting phase boundary[8]. The reason for such inconsistency is still unresolved, and is possibly subject to the diverse sensitivity timescales of the different probes, the slight variation of the oxygen content, and field-induced effect.

**Focus**

We performed a systematic electric transport study on optimally Ce-doped $La_{2-x}Ce_xCuO_{4\pm\delta}$ (LCCO, $x$ = 0.1) thin films with finely-tuned oxygen composition. The superconductivity in these films can be suppressed in fields below 10 Tesla. The present investigation focuses on the electronic behavior of the normal state in extremely high magnetic field up to 58 T. The combined experiments on electrical resistivity, $\rho_{xx}(B)$, Hall resistivity, $\rho_{xy}(B)$, and Hall coefficient, $R_H$, result in an exceptional multidimensional phase diagram ($T, B, \delta, x$) constructed in conjunction with previous studies of the Ce-doped phase. This work unveils remarkable key features which suggest the formation of conductive nano-filaments (CNF) co-exising with the AFM. Such conductive network explains the nontrivial AF behaviour and the diversity of all existing experimental data, collected in antiferromagnetic, superconducting and normal states.

The CNF may arise due to polarisation of neighbouring oxygen orbitals analogous to a spin-orbital polaron. It is created due to strong electron-electron correlations or strong on-Cu-site Coulomb repulsion, resulting from a competition between the polarisation of the oxygen orbitals and the kinetic energy of the self-trapped electron (see, SI2, Eqs. (5,6,21) and Figures S4,S8-10, where the electronic structure of SO polarons and CNFs are given for details). As a result of the Coulomb repulsion many electrons may become confined into

CNFs, which can orient along two perpendicular directions. Similar structures formed from conventional[25] and spin-bag polarons[26] have been noticed earlier for other systems.

**Crystal films growth and oxygen doping**

The LCCO thin films were grown on (001)-oriented $SrTiO_3$ substrates by pulsed laser deposition (PLD). To achieve the slight variation in the oxygen content, the samples were subjected to a controlled annealing process. The stoichiometric composition of the material was either under-annealed (UD), optimally-annealed (OP) or over-annealed (OD). The optimally annealed samples (OP1) show the highest $T_c$, while both the under-annealed (UD1) and over-annealed (OD1) samples exhibit slightly lower $T_c$ (Fig.1, g-i). The lattice structure of all the samples, subject to different annealing times, was meticulously ascertained by X-ray diffraction. The present $\varphi$ scans (Fig.1, a-c) and reciprocal space mappings (Fig.1, d-f) demonstrate high quality epitaxial growth films, with no impurity phases seen from the θ/2θ scans (Fig.S1 in SI1).

It is difficult to establish exactly the slight oxygen variation in the thin films. A more meaningful approach would be to measure the c-axis lattice parameter, as it reflects the change in the oxygen content qualitatively [30]. Progressing from under- to over-annealed samples, the c- lattice constant gradually shrinks, showing a positive correlation with oxygen deficiency. In Fig 2a, the full superconducting transition temperature $T_{c0}$ is plotted as a function of the c-axis lattice constant. Notably, the oxygen dependence of the superconducting transition, $T_{c0}(\delta)$, exhibits a similar dome-like behaviour as was observed for $T_{c0}(x)$ variation with Ce doping.

**Electrical transport**

Longitudinal $\rho_{xx}(T)$ resistivity was measured as a function of temperature on the annealed LCCO thin films. The behaviour in $\rho_{xx}(T)$ is largely metallic for all samples down to $T_c$ in zero field. However, the under-annealed sample UD1 shows a minute upturn in $\rho_{xx}(T)$ just prior to $T_c$, comparable to the behaviour observed in underdoped LCCO with $x < 0.10$ (Fig 1c). In a magnetic field of 15T an upturn develops in $\rho$ for all compounds, characteristically right at (or below) the zero-field superconducting transition temperature (Fig 1c). The significant reversal in the overall temperature dependence of $\rho_{xx}(T)$ below $T_c$ between the 0T and 15T states suggests anomalous magnetic field dynamics in these systems. Similar properties in the normal state may be reminiscent of strong spin-fluctuations.

**Hall resistivity and Hall coefficient**

The Hall resistivity $\rho_{xy}$ allows the determination of temperature and field dependence of the Hall coefficient $R_H$ in the annealed LCCO films. Focusing on the normal state results in 15T, the Hall coefficient is negative for all samples suggesting electrons as dominant charge carriers. Additionally, $R_H$ shows unusual temperature dependence with a slight (nearly flat) negative curvature with decreasing temperature up until T ∼ 25K. Below this temperature scale T < 25K the Hall coefficient decreases sharply with a positive correlation (Fig 2d). The positive dependence at low temperatures is particularly strong in under-annealed (UD) and optimally-annealed (OP) films. In fact the behaviour of $R_H$ in under- and optimally- annealed samples is near indistinguishable for T ∼ 25K. The effect is substantially less pronounced in over-annealed (OD) samples.

The non-constancy of $R_H$ as a function of temperature is quite surprising and suggests significant changes to either carrier or scattering mechanisms in LCCO below T < 25K. The sharp decrease of $R_H$ could be consistent with multiple-relaxation timescales[50], spin/charge density wave formation[48], or an enhancement in spin-orbit coupling.[46] The pronounced correspondence of the anomalous temperature dependence in $R_H$ to under-annealed and optimally-annealed LCCO postulates a link between the origin of this unconventional behaviour and antiferromagnetism, leading towards superconductivity.

The Hall coefficient $R_H$ similarly demonstrates unconventional field dependence. For temperatures below T < 25K $R_H$ develops a pronounced broad minimum at ~20T (Fig 2b). The minimum shows slight temperature dependence and emerges only in the normal state of the LCCO. For temperatures T ≥ 25K the value of $R_H$ drops significantly (by ~75%) and exhibits a nearly flat correlation in field. Such unusual field scalability of the Hall coefficient implies markedly different transport processes at low temperatures T < 25K.

Consolidating the behaviour of the Hall coefficient with temperature and field identifies two characteristic temperature scales $T_1$ = 62.5 ± 7.5K and $T_2$ = 25 ± 5K (Fig 2c). The first temperature crossover $T_1$ is insensitive to the oxygen doping level, and remains constant within experimental error up to a maximum measured field of 58T. The second temperature scale $T_2$ shows more variation to both the oxygen deficiency and magnetic field. The $T_2$ in under-annealed (UD) and optimally-annealed (OP) samples displays shallow broad field dependence. This may be consistent with substantial changes in the scattering processes, mobility or charge carrier density.

**Magnetoresistance and anomalous field dependence**

The longitudinal $\rho_{xx}(B)$ and Hall $\rho_{xy}(B)$ electrical resistivities in optimally-annealed samples (OP) were studied as a function of magnetic fields up to 58T at low temperatures T ≤ 26K (Fig 3a & 3b). The Hall resistivity $\rho_{xy}(B)$ shows a substantial broad minimum in field over all reported temperatures (Fig 3a). The minimum is most pronounced at the lowest temperature of 4.2K. The exact position of the minima itself shows intricate field and temperature, as well as sample dependence (Fig 3c).

The position of the minima $B_{min}(T)$ in $\rho_{xy}(B)$ of two optimally-annealed samples (OP1 and OP2) was traced through a range of temperatures 4.2 < T < 26K (Fig 3c). The minima show weak negative or nearly flat dependence until a temperature of T ∼ 15K (for OP1) and T ∼ 17K (for OP2), followed by a more drastic positive slope at higher temperatures. The data suggest a characteristic temperature scale $T_{min}$, which is sample dependent, associated with a full reversal of magnetoresistive properties in optimally-annealed LCCO. Plotting the temperature scale $T_{min}$ as a function of superconducting transition temperature $T_{c0}$ shows super-linear correlation (Fig 3f). Similar behaviour switching in magnetoresistance from positive to negative with field may be expected in systems with 2D spin glass formation, and/or transitions from weak to strong localization. [47]

Conversely, the longitudinal resistivity $\rho_{xx}(B)$ shows a maximum in its field dependence at low temperatures (Fig 3b). The maximum is most evident at the lowest measured temperature of 4.2K. The position of the maxima $B_{max}(T)$ is followed precisely through a range of temperatures for two optimally-annealed samples (OP1 and OP2) (Fig 3d). The variation in the minima location demonstrates strong temperature dependence with a turning point at T ~ 20K (for OP1) and T ~ 17K (for OP2). Extracting the characteristic scale $T_{max}$, which shows moderate sample dependence, and plotting it against the superconducting transitions temperature $T_{c0}$ presents a convincing linear correlation (Fig 3f). The sharp contrast in behaviour between the longitudinal and Hall resistivities may be indicative of field-enhanced spin-orbit coupling, screened magnetic scattering and electron confinement to 2D charge sheets consistent with a CNF scenario. [49]

**Discussion**

To summarize the previous works on LCCO all transport probes [6, 23, 24] point to a static antiferromagnetic (AF) order beyond $x$ = 0.10, while the zero-field low energy $\mu$SR only detects static magnetism below $x$ = 0.08 [8]. Understanding the fundamental physics beyond this discrepancy [6–8] that occurs on the boundary of the three-dimensional AFM relies on precision high magnetic field experiments in LCCO samples with finely controlled oxygen content – as performed in the present study.

Several viewpoints exist regarding this unresolved conundrum. It has been predicted that the competition between spin density wave (SDW) order and superconductivity shifts the quantum critical point (QCP) to lower doping levels, and the QCP reverts back when the superconductivity is destroyed by magnetic field [33]. Additionally, a double QCP scenario has been proposed where the two QPCs at the boundaries of antiferromagnetism[8] and Fermi liquid[7], move towards opposite directions in fields, see in the top inset of Fig. 4. It is also interesting to note that in very thin films of electron-doped cuprates one of the QCPs is associated with two-dimensional superconductor-insulator quantum phase transition determined through ionic liquid gating studies [34, 35] (see, also, for a comparison Refs [36–38]). This published work concludes that the quantum phase transition: 1) has a percolation character, where the (super) conducting state may percolate via the formation of a connected electron spider web with electron conduction along it; 2) the resistance dependence on magnetic field B tends to saturate at high B and the magnetoresistance vanishes.

The present results demonstrate non-linear and non-monotonic longitudinal and Hall resistivities in a magnetic field, strongly temperature and field dependent Hall coefficient, vanishing magnetoresistance at high fields with several dynamic temperature- and field-scales $T_1$, $T_2$, $B_{max}$ and $B_{min}$ respectively. Our current findings may unequivocally corroborate all the proposed scenarios into one coherent picture and offer a consistent interpretation for all reported experimental observations.

**Magnetotransport and spin polarons**

The Hall resistivity $\rho_{xy}(B)$ is proportional to B at $T > T_1$, a feature that can be attributed to a simplified single band or a compensated two bands picture in metals[31, 32]. Below $T_1$ = 62.5 ± 7.5 K, $\rho_{xy}(B)$ is no longer linear, and the Hall coefficient $R_H$ is no longer constant (Fig. 2b). This very unusual behaviour of the Hall resistivity may be associated with spin-orbital (SO) polarons, which self-organize into quasi-one-dimensional islands (strings) and threads that collectively induce the observed changes in $R_H$. This can be explained due to a comparatively small kinetic energy of the SO polarons with respect to their Coulomb interaction. The characteristic temperature $T_1$ is independent of fields up to 58 T, and is defined by the competition between the kinetic energy of SO polarons and their Coulomb many-body interaction. Present results imply a characteristic value of the inter-polaron Coulomb interaction of the order of $T_1$ [8, 24].

Approaching the second temperature scale $T_2$, $\rho_{xy}(B)$ changes more dramatically, and manifests a kink in the Hall coefficient $R_H(T)$ at $T_2$ = 25 ± 5 K (Fig. 2d), roughly following the temperature for the upturn in resistivity $\rho_{xx}$. This upturn may reflect the appearance of Anderson localization arising due to the low-dimensional character of the conductive nano-filament (CNF) network. This suggests tentative freezing of electrons, due to creation of a spin or charge-density wave gap in the electronic spectrum, leading to a loss of conduction carriers [22]. The second critical temperature found in the present work, $T_2(B, \delta)$ is not as stable as the $T_1(B, \delta)$, and is mainly attributed to a much weaker inter-plane exchange coupling ($J_\perp$) of the order of 1-2 meV, indicative of 3D antiferromagnetic interactions. A self-consistent multidimentional phase diagram is established that signifies that $T_1$ and $T_2$ in ($T, \delta, x$ = 0.10) intersect with the ($T, \delta = 0, x$) at the starting points of twofold in-plane anisotropic magnetoresistance (AMR) and resistivity upturn, corresponding to 2D and 3D AFM, respectively (Fig. 4).

At temperatures $T < T_2$, the developing maxima and minima in $\rho_{xx}(B)$ (at $B_{max}$) and $\rho_{xy}(B)$ (at $B_{min}$) respectively at high fields suggest field-stimulated electron transport mechanism. In particular we found that such transport can be associated with CNFs and SO polarons, which are low-dimensional structures. In magnetic field both may become polarised which decreases scattering. Therefore, the samples containing polarised CNFs demonstrate greater conductivity (see, the sections I-IV from SI2). Similar effects of vanishing magnetoresistance have been observed in conducting one-dimensional chains of polymers, in polyacetylene nano-fibers, [27, 28] underlining the importance of Coulomb interactions between polarons[29], comparable to the present case.

Additionally, the dip in the dependence of $R_H(B)$ arising at $T < 25K$ (see, Fig. 2b) may be related to a decoupling of electronic strings from the rest of the conducting structure forming quasi-one-dimensional channels. At long distances the polarons are repelling each other due to poor screening of the Coulomb forces. While at short distances they prefer to establish a common quasi-one-dimensional polarisation well, where the features of individual polarons are washed out and a string [25] or confined Luttinger liquid is formed.

**Spin-density wave, superconductivity and Luttinger liquid**

Both $B_{max}(T)$ and $B_{min}(T)$ curves display non-monotonic behaviors with temperature (Fig 3). The two limiting temperatures $T_{max}$ and $T_{min}$ where the $B_{max}(T)$ or $B_{min}(T)$ reach an extremum display a positive correlation to $T_c(\delta)$, implying a link between the spin density wave (SDW) fluctuations and superconductivity. Consequently, the two characteristic temperatures $T_1$ and $T_2$ are naturally associated with two- and three-dimensional AFM respectively. Such transformation of the AF order from 3D to 2D is correlated with the existence of CNF network created by the doping into the AF state (see Fig. 4 right inset, details in SI2).

Within the proposed CNF network the formation of one-dimensional channels leads to a unique stabilisation of Luttinger liquid at temperature below $T_2$. Inside the channels the electron and spin degrees of freedom are completely decoupled (see, Fig.5, filament band). In order to perform a tunnelling transition between two neighbouring parallel channels a charge should be combined with a spin or it may hop only as an electron pair (see, Fig. 6). The nuance here is that in the one-dimensional Hubbard model single-particle hopping is irrelevant, and that weak interchain coupling doesn't cause coherent particle motion in the transverse direction [43].The coherence can only be restored when the electrons are tunnelled in pairs [44]. And such tunnelling pairs can be condensed and induce superconductivity while in a normal state the system may behave as 2D Luttinger liquid [45]. This unconventional tunnelling mechanism has been originally proposed for inter-plane tunnelling in cuprates[43]. Thus, the CNF network might play a key role in the mechanism for superconductivity in cuprates.

## CNF parquet model and antiferromagnetism

To understand the behaviour, $\rho_{xx}(B)$ and $\rho_{xy}(B)$ in magnetic field $B$ we have considered a theoretical model[39], which can be applied to a two component hetero-system. Specifically, we have chosen a "parquet" model where one phase consists of quasi-one dimensional filaments – the CNF state which is penetrating or depleting the AF state, while the other phase is the 2D AF plane being depleted in turn (see, [39] and SI2). The two stacked nano-structures form a characteristic "parquet" shape (see, right insert on Fig. 4). The superconductivity has been stripped away in the considered calculations and only contributions from the normal state are evaluated. The magneto-conductivity for the proposed model was calculated by methods based on dual and conformal transformations (see for details [39]). Notably, the parquet hetero-structure consists of a semiconducting AFM and a conductive CNF framework. The current carriers in the AF state have short scattering time associated with spin fluctuations and a large effective mass, while in the conducting one-dimensional channels creating the CNF the scattering time is long and the effective mass is small.

Using the "parquet" model we were able to successfully simulate both $\rho_{xx}(B)$ and $\rho_{xy}(B)$. Indeed the dependence $\rho_{xx}(B)$ shows a positive magnetoresistance at small field and negative magnetoresistance at large field. Simultaneously, the Hall resistivity shows a minimum at $B_{min}$ (see, Fig. S11 and Fig. S12 in SI2) as observed in the present experiments (see, Fig. 3 a,b). Thus the theoretical calculations based on the "parquet" model show good qualitative and semi-quantitative agreement with the existing data (see, Fig. S11 and Fig. S12 in SI2 where the theoretical dependencies for $\rho_{xx}(B)$ and $\rho_{xy}(B)$ are calculated for typical parameters of a two phase system, consisting of AF plane and the CNF).

## CNF and superconductivity

An important finding is that both $B_{min}(T)$ and $B_{max}(T)$ show a positive relation to $T_{c0}$ (Fig. 3e and 3f), suggesting a significant correlation between SDW fluctuations and superconductivity. We realize that for spin-bag polarons a relation between the superconducting, $\Delta_{SC}$ and the SDW gap, $\Delta_{SDW}$, has been given previously in the form $\Delta_{SC} = \Delta_{SDW} \exp[-1/N(E_{k_F})U]$ where $N(E_{k_F})$ is the density of states at the Fermi energy and $U$ is the pairing interaction[26]. Although SO polarons have a different origin, i.e. they arise due to a polarisation of oxygen orbitals created by strong correlations or a large Hubbard $U$, they can also provide a similar relation between $\Delta_{SC}$ and $\Delta_{SDW}$ (see, SI2, for details).

When the AF interaction between $CuO_2$ layers is weakening, the 3D AF order vanishes while only 2D AF order remains. Such a transition from 3D to 2D AF order happens when $T > T_2$ and shifts slightly with tuning of the oxygen content. At these temperatures, due to the CNF formation, only short-range order survives in the plane. The competing interactions induced by this change can also be responsible for the observed non-trivial behaviour of the second QCP arising at doping level $x_{FS}$ when the AF spin liquid state and CNF phase vanish while the SC state still remains. The metamorphosis of the described AF states identified in the analysis of the present experiments highlights the complex nature of the coexistence of magnetism and superconductivity, when they are in a position of strong competition and when the CNF network percolates and depletes the AF order.

Present findings and the underlying theoretical interpretation may offer a natural explanation to the recent ground-breaking discoveries by Bozovic et al that the superfluid density in overdoped cuprates is directly proportional to $T_c$ (see SI3).[52] In CNF framework the superfluid density is proportional to the number of paired tunnelling processes between the nano-filaments. As the nano-filaments percolate into a 3D web the number of tunnelling transitions decreases and so does the superfluid density (SI2, Section IV and SI3).

**SUMMARY**
Here we describe how superconductivity emerges in cuprates when electrons are doped into their antiferromagnetic, insulating CuO plane. We demonstrate evidences that these electrons support spin fluctuations that may be key to the superconductivity and a mysterious QCP intervening inside the superconductor. We also show that these electrons are polarizing oxygen orbitals. Such local polarization traps the electrons and transforms them into cigar-shaped spin-orbital (SO) polarons. Together they self-organize into conducting a nano-filament (CNF) structure, which depletes the antiferromagnetic insulator, changing its order from 3D to 2D and inducing spin-fluctuations, spin gap and superconductivity. At higher doping the CNF state may transform to a nematic electron liquid phase, leading to the enhancement of the superconducting transition temperature $T_C$ via the transverse motion of the SO polarons, similar to transverse stripe fluctuations.[51] We construct a possible relation between them based on all the results presented in the form of a holographic phase diagram.

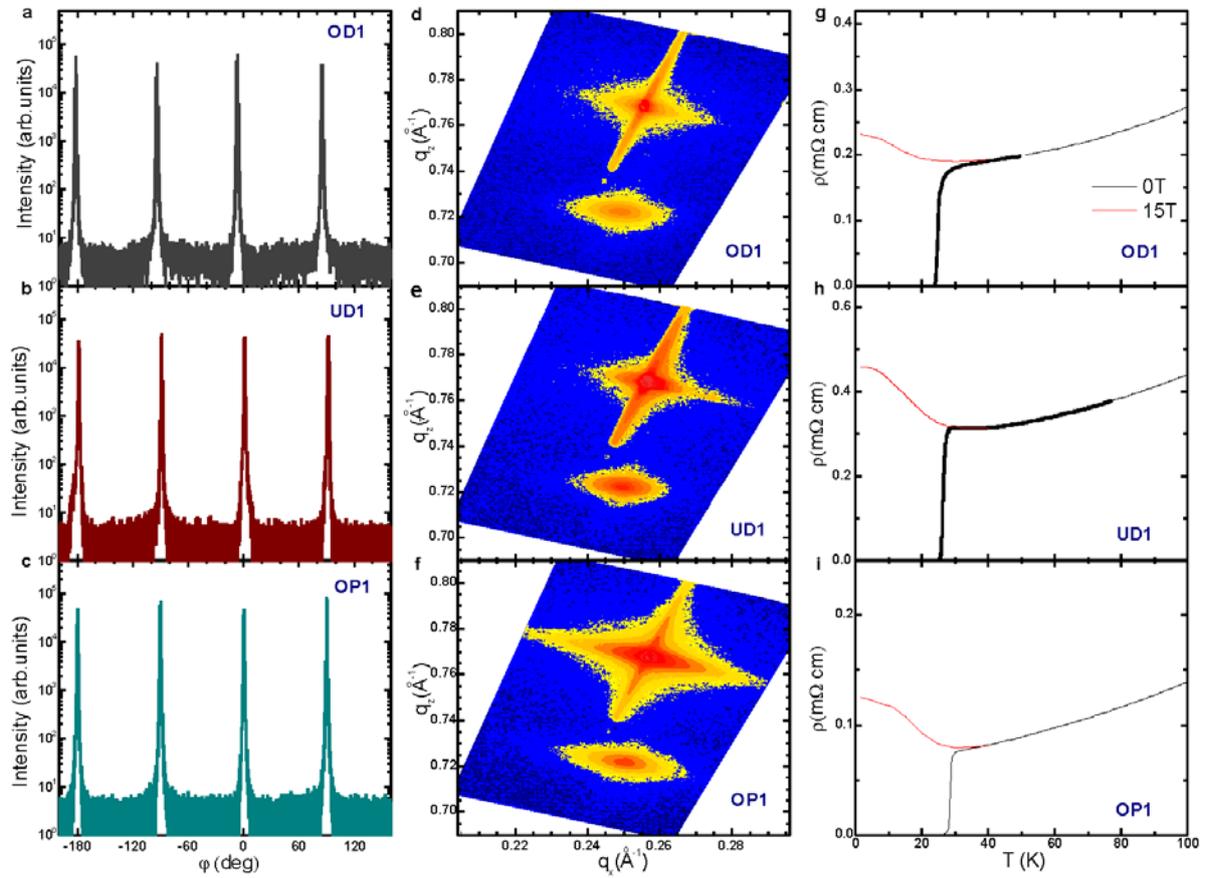

**Figure 1**: Structure characterizations and resistivity for LCCO ($x$ = 0.1) thin films. **a-c**, The $\varphi$ scans of (103) plane for samples UD1, OP1 and OD1. There are four peaks of nearly equal height, reflecting the high quality of the samples. **d-f**, The reciprocal space mapping of (103) plane of SrTiO$_3$ (cross red pattern) as well as (109) plane of LCCO (oval red pattern). **g-I**, Temperature dependence of the resistivity at 0T and 15T with B∥c. An upturn in $\rho(T)$ is observed at a characteristic temperature scale $T_2$ in the normal state.

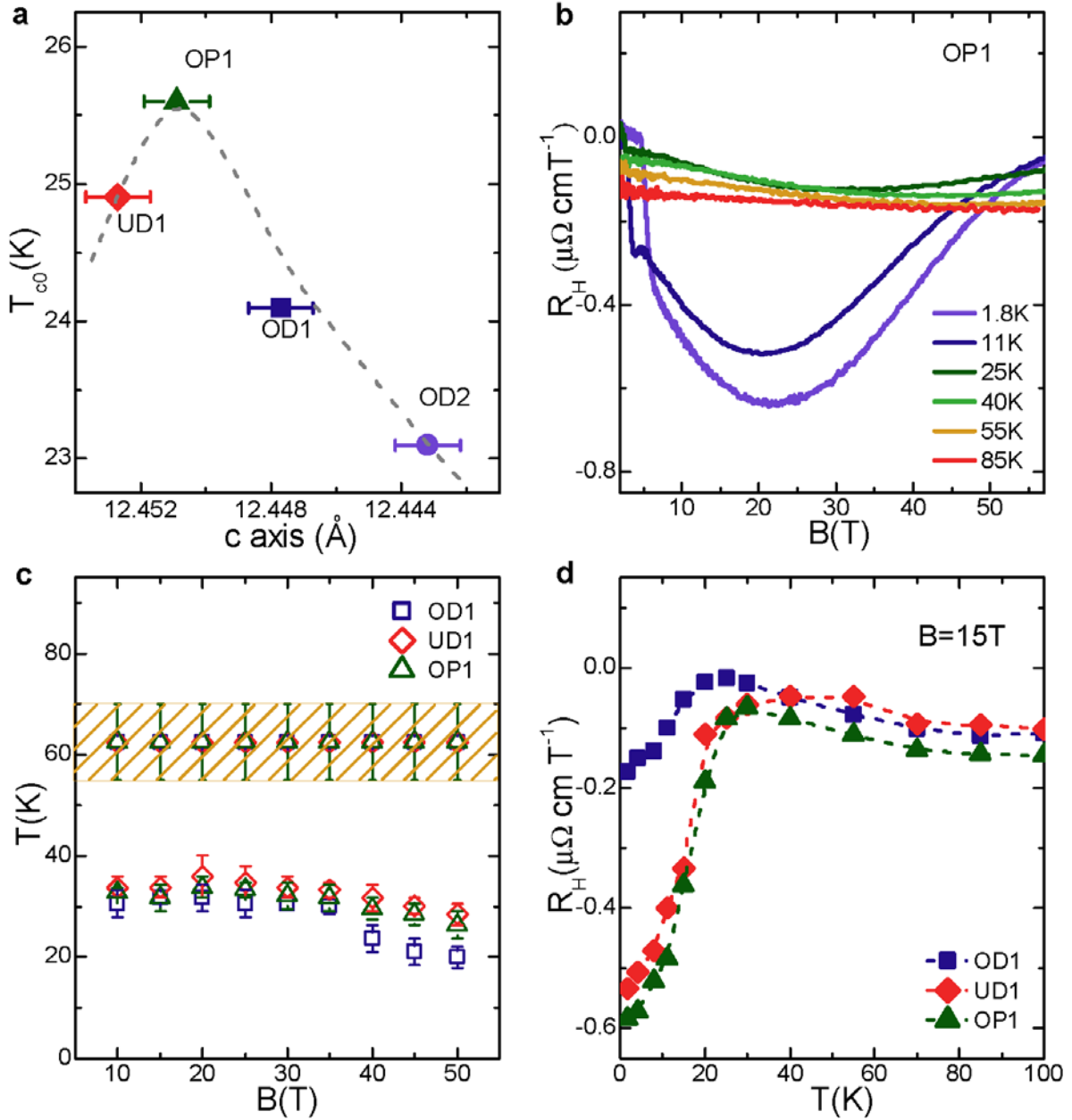

**Figure 2**: Lattice parameters and Hall signal in field. **a**, Relation between $T_{c0}$ and c-axis lattice parameter. The $T_{c0}(c)$ shows dome-like behaviour, similar to the case observed with Ce-doping. **b**, Magnetic field dependence of the Hall coefficient $R_H$ from 1.8 K to 85 K. The Hall coefficient is almost constant with increasing field at 85 K, which reflects that the Hall resistivity $\rho_{xy}(B)$ is proportional to the magnetic field. But at temperatures below $T_1 = 62.5 \pm 7.5$K, the Hall coefficient $R_H$ is no longer constant, as $\rho_{xy}(B)$ starts to develop a curvature. The deviation from linearity is particularly evident in the Hall coefficient for temperatures below $T_2 = 25 \pm 5$K. **c**, Two characteristic temperature scales in electron-doped $La_{2-x}Ce_xCuO_{4\pm\delta}$ thin films: $T_1$ at $62.5 \pm 7.5$K is robust in magnetic field for all samples (squares for OD, diamonds for UD and triangles for OP); $T_2$ at $25 \pm 5$K, marking the upturn in $\rho(T)$ in the normal state, shows field dependence above 30T. **d**, Temperature dependence of the Hall coefficient $R_H$ shows a kink between 20 K and 30 K, roughly trailing the $T_2$.

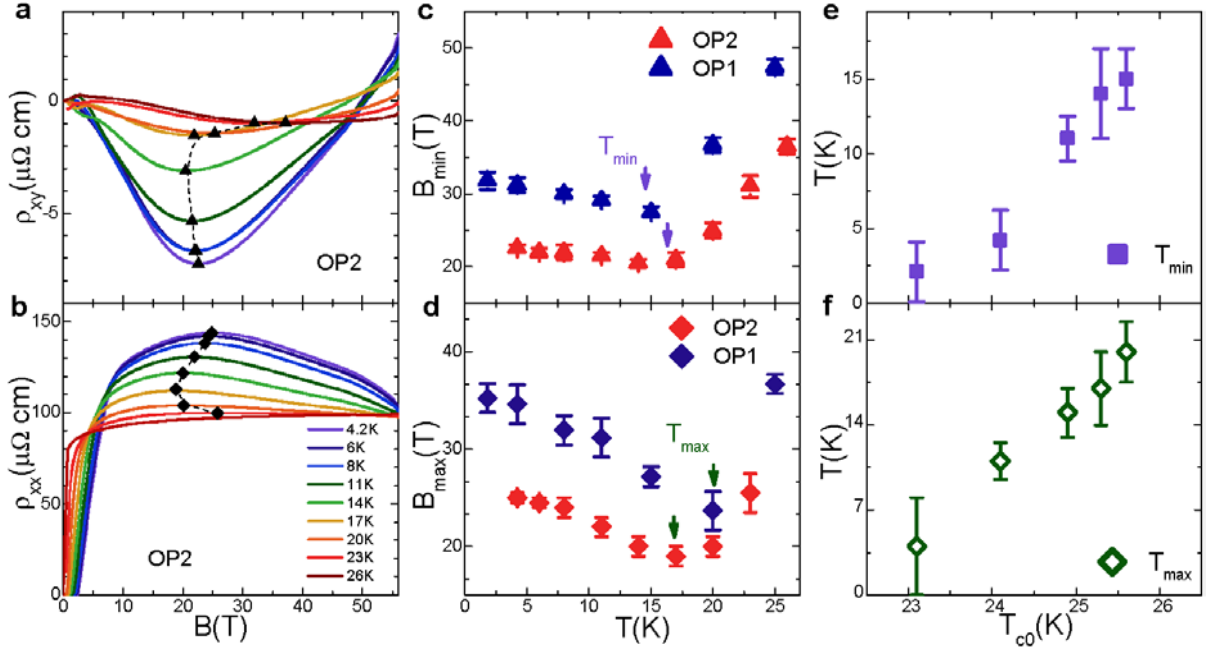

**Figure 3**: Correlation between spin density wave (SDW) and superconductivity. **a**, $\rho_{xy}(B)$ shows a minimum from 4.2 K to 26 K that becomes more pronounced with decreasing temperature. The field location of the minimum is denoted by $B_{min}$. The dash line is a guide to the eye. **b**, $\rho_{xx}(B)$ exhibits positive followed by negative magnetoresistance (MR) as the field increases, with the effect heightened as the temperature is decreased from 26 K to 4.2 K. The dash line separates the positive and negative MR regimes. The magnetic field corresponding to a maximum in $\rho_{xx}(B)$ is marked as $B_{max}$. **c**, The temperature dependence of $B_{min}$ for two optimized samples (OP1 and OP2). As a function of temperature, $B_{min}(T)$ first gradually decreases and then increases rapidly, suggestive of a closed SDW gap and enhanced fluctuations. The temperature where $B_{min}(T)$ shows a minimum is marked as $T_{min}$. **d**, The temperature dependence of $B_{max}$ for OP1 and OP2. The temperature where $B_{max}(T)$ shows a minimum is marked as $T_{max}$. **e** and **f**, $T_{min}$ and $T_{max}$ display a positive relation to the superconductive transition temperature $T_{c0}$ for the five samples UD1, OP1, OP2, OD1 and OD2. The behaviour of $\rho_{xx}(B)$ and $\rho_{xy}(B)$) is very well described by the theoretical "parquet" model, see Fig. S11 and Fig. S12 in SI2.

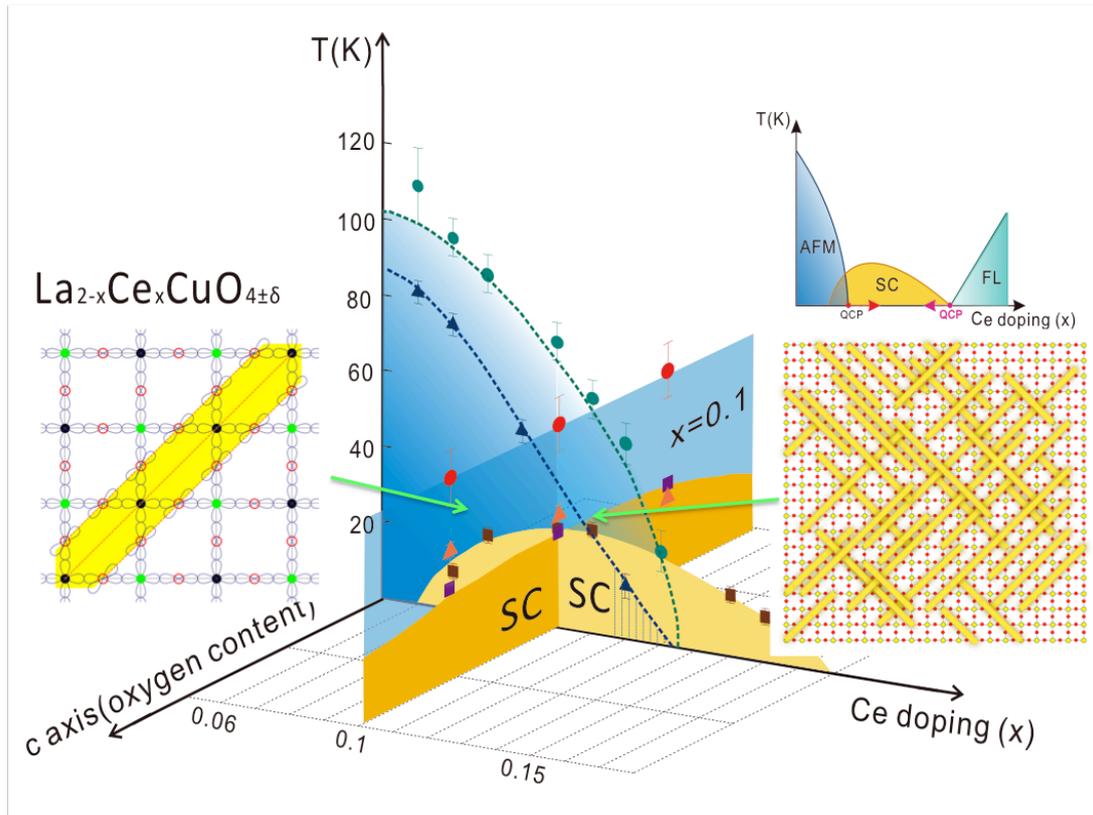

**Figure 4**: A multidimensional phase diagram of LCCO as a function of Ce and oxygen content. Along the Ce doping axis, the AFM boundary (blue area) is determined by transport measurement in a magnetic field. Specifically, the border of the two-dimensional AFM regime is established by in-plane angular magnetoresistance measurements (green spheres). This region exhibits strong electron-electron correlations at small doping levels, so that spin-orbital (SO) polarons can arise. A SO polaron is created when a doped electron polarizes free (non-bonded with Cu) p-orbitals of the oxygen atoms nearest to the few copper sites that have the same spin orientation. The polarized orbitals are shown in yellow (the left inset). This type of polarization leads to orbital hybridization that gives rise to the next-nearest neighbour hopping originated via the electron path, Cu-O-O-Cu. As a result the doped electron becomes self-trapped by the orbital polarization potential which has a cigar shape schematically presented by yellow region on the left inset (see, Figures S2,S3 in SI2 for details). With doping these SO polarons self-organize into conducting nano-filaments(CNF), which form a quasi-one-dimensional percolating electron network – a spider web, schematically shown in the right inset (bottom). The red dots there represent the copper sites, forming the square lattice in the $CuO_2$ plane, the green/red solid bold lines represent the CNFs, with spin up/down orientation. Due to the quasi-one-dimensional character of this electron network Anderson localization can arise at temperatures below $T_2$ (blue

triangles). This phenomenon may correspond to the upturn in the normal state resistivity ρ(T) and the kink in the temperature dependence of the Hall coefficient $R_H$ observed between 20K and 30K (see, Fig. 2) which is roughly following $T_2$. Along the c-axis (oxygen doping), for the optimal Ce-doping LCCO (x = 0.1), the 2D AFM boundary ($T_1$, red spheres) is quite robust against oxygen content. Above $T_1$, the electron network evaporates into the SO polaron gas. However, the edge of the 3D AFM, manifested as a kink in $R_H(T)$ curves (red triangles), shifts slightly with tuning of the oxygen content. The top right inset is a sketch of a Ce-dependent phase diagram in zero field. There are two potential quantum critical points (QCPs) located in proximities of AF[8], and Fermi Liquid[7] phases, respectively. It has been suggested that the QCPs move in opposite directions with magnetic field (indicated by arrows) when the superconducting state is suppressed. Such behaviour of the QCPs can be explained by the existence of CNF web, as indicated by present experiments.

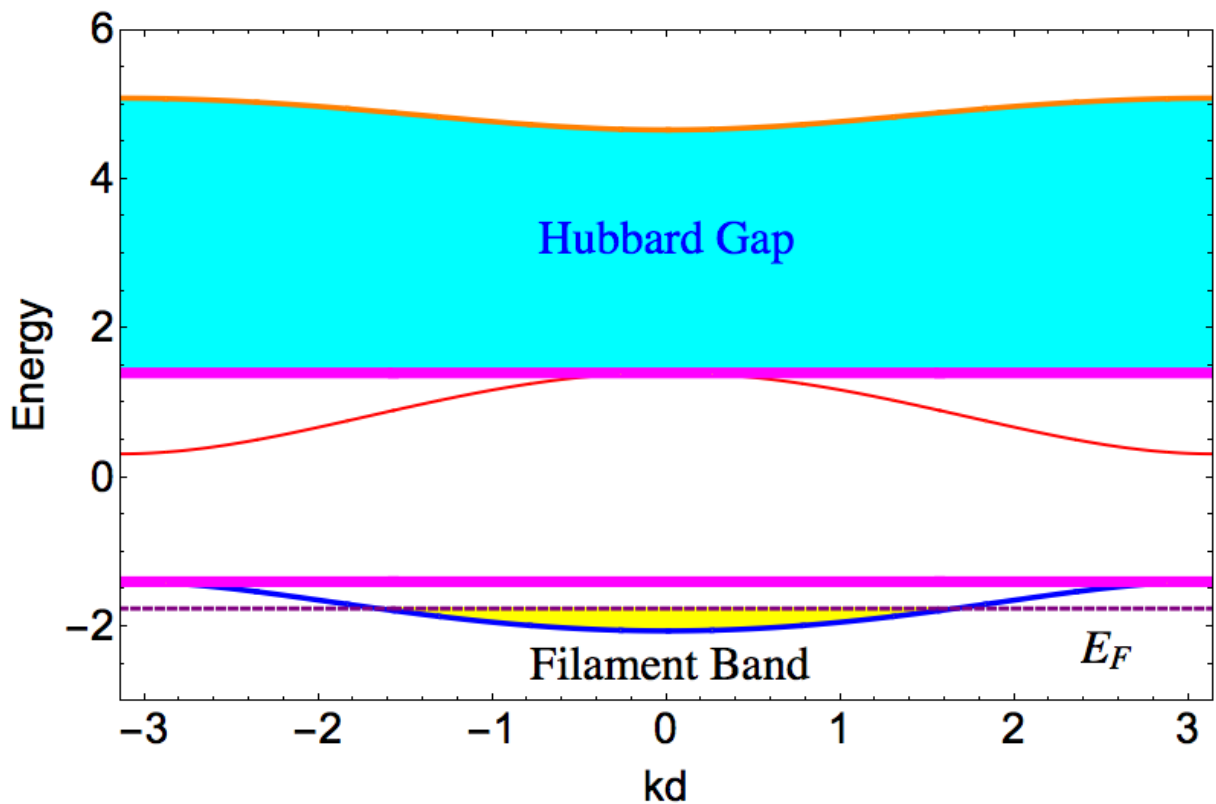

**Figure 5**: The electronic band structure of an individual nano-filament. With doping electrons are filling the lowest band. The position of the Fermi energy $E_F$ is noted by the dashed purple line. A low-dimensional Luttinger liquid state arises below the Fermi energy (noted in yellow). At the Brillouin Zone boundaries the top of the lowest band merges with a flat band (shown in magenta). The electron density of states (DOS) increases significantly at the zone edges, which may strongly enhance superconducting pairing. There are two such flat bands whose positions are determined by the polarisation of the oxygen orbitals only. The lowest flat band corresponds to the energy limit $E=-p$, while the upper one is linked with the energy $E=p$, where $p$ is a hopping integral between oxygen sites associated with the polarisation of its orbitals. The Hubbard gap (shown in cyan), separates the lowest four oxygen bands from the upper Hubbard band. Its value is proportional to the constant of the Hubbard interaction, $U$. It is most remarkable that the position of the lowest filament band is independent of the value of $U$ which is highly reminiscent of the formation of a Zhang-Rice singlet in the hole-doped cuprates. Here the energy is expressed in the units of $t$ – the hoping integral between Cu and O sites. In this example we have taken $p=1.4\ t$ and $U=4t$.

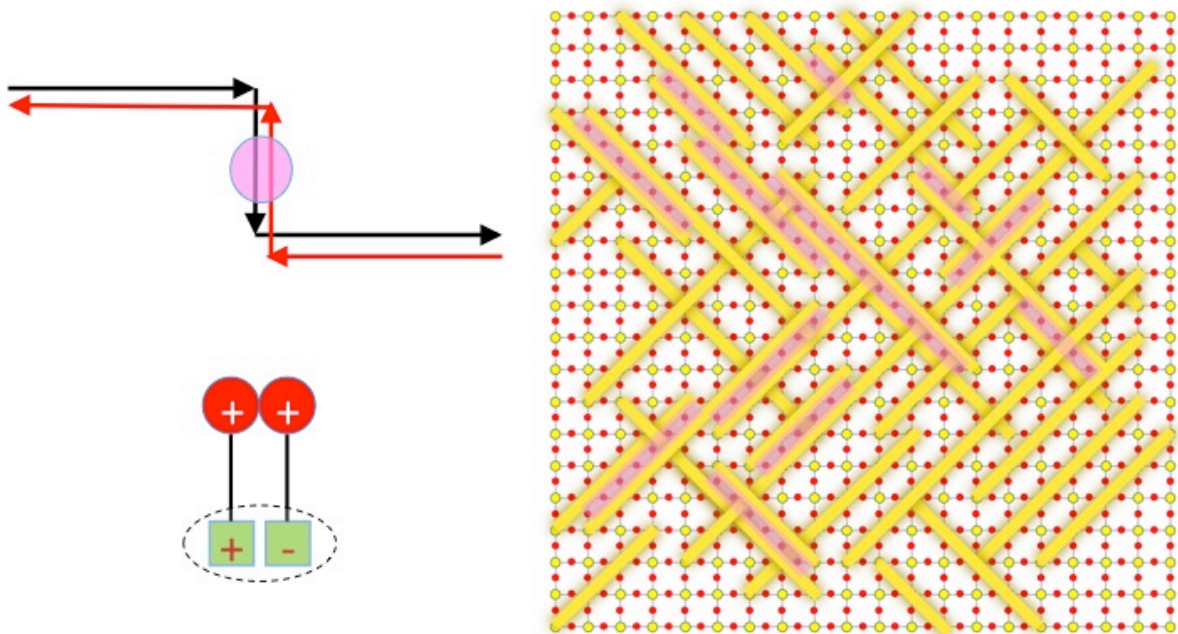

**Figure 6**: The superconducting pairing relation to the CNF state. The superconducting pairing may be associated with a two-electron inter-chain tunnelling across two neighbouring, parallel CNFs (top left – spin up electron tunnelling (red) with spin down electron tunnelling (black) to form a confined vortex (pink)). The vortex confinement is equivalent to the formation of vortex-particle composites (bottom left – vortexes (green) where the negative sign denotes the antivortex, particles (red) with the charge indicated, for details see the supplementary file SI3). In this specific example (bottom left) the confinement corresponds to the formation of a bound vortex-antivortex pair with a total electric charge equal to +2. This unusual decoupling mechanism corresponds to a vortex-antivortex unbinding, arising via the Kosterlitz-Thouless phase transition. Therefore the superconducting critical temperature $T_c$ becomes naturally proportional to a superfluid density $n_s$, as observed in the Ref.[52].

## Acknowledgements

We would like to thank R.L. Greene, L. Shan, S.L. Li, Z.Y. Meng for fruitful discussions. H.Y thanks L.H. Yang for assistance in structural characterizations. This research was supported by the National Key Basic Research Program of China (2015CB921000), the National Natural Science Foundation of China Grant (11474338), and the Strategic Priority Research Program (B) of the Chinese Academy of Sciences (XDB07020100).

## Author contribution

H.Y. prepared samples and performed structural characterizations. H.Y., G.H. Z.L. and J.W did transport measurements. All the authors contributed to the discussions and writing; K.J. supervised the project.

## Competing financial interests

The authors declare no competing financial interests.

Correspondence and requests for materials should be addressed to J.W. (email: jfwang@hust.edu.ac) or to F.V.K. (email: F.Kusmartsev@lboro.ac.uk) or to K.J. (email: kuijin@iphy.ac.cn).